\documentstyle[amssymb,12pt,manuscript,aps,epsfig]{revtex}
\begin{document}
\title{Exclusive  $\phi$ production in proton-proton collisions 
in the resonance model.}
\author{Amand Faessler$^a$, C. Fuchs$^a$, M.I. Krivoruchenko$^{b}$, 
 B.V. Martemyanov$^{a,b}$.}
\address{$^a$ Institut f\"ur Theoretische Physik der Universit\"at T\"ubingen, 
Auf der Morgenstelle 14, D-72076 T\"ubingen, Germany}
\address{$^b$ Institute for Theoretical and Experimental
Physics, B. Cheremushkinskaya 25, 117259 Moscow, Russia}
\maketitle  
\begin{abstract}
The exclusive $\phi$ meson production 
in proton-proton reactions is calculated within the resonance model.
The considered model was already successfully applied to the
description of $\pi$, $\eta$, $\rho$, $\omega$, $\pi\pi$ production in
proton-proton collisions. The only new parameter entering into
 the model is the $\omega-\phi$
mixing angle $\theta_{mix}$ which is taken equal to 
$\theta_{mix} \approx 3.7^o$.
\end{abstract}

\pacs{13.60.Le, 14.20.Gk, 25.40-h}
\section{Introduction}

The production of $\phi$ mesons is strongly suppressed compared to that of
$\omega$ mesons. This fact is known under the name ``OZI rule''\cite{OZI}.
According to the OZI rule $\phi$ mesons can only be produced 
due to a small admixture of non-strange light quarks in their wave
function. The corresponding mixing angle $\theta_{mix}$
is equal to $\theta_{mix} \approx 3.7^o$\cite{PDG}. According to this 
mixing angle  the ratio of
$\phi$ and $\omega$ mesons cross sections should at comparable
energies naively be equal to
\begin{equation}
R_{\phi/\omega} = \tan^2 \theta_{mix}\cdot F \approx 4.2\times 10^{-3}\cdot F~,
\label{naive}
\end{equation}
where $F$ is a correction due to the different phase space factors 
for $\phi$ and $\omega$ mesons.
Experimentally, the ratio $R_{\phi/\omega}$ is in $pp\rightarrow
pp\phi(\omega)$  reactions, however, known to be one order of 
magnitude larger than the naive expectation \cite{DISTO,Blo75,Bal77,Are82,Gol97}.
Various models have been used as a framework for theoretical
attempts to explain this large discrepancy, ranging from simple 
one-pion exchange models \cite{Sib96} to more sophisticated effective Lagrangian 
models \cite{Tsushima03}. Also the  resonance model is successfully 
explaining the production of a large variety of mesons such as 
$\pi$, $\eta$, $\rho$, $\omega$, $\pi\pi$ 
in elementary nucleon-nucleon reactions \cite{Teis,Fuchs03}. 
In the present note we extend the application of the resonance model 
to the description of $\phi$ meson production
in $pp$ collisions. In the next section we briefly describe 
the resonance model and present
the results for the $\phi$ meson production. 
Then we discuss the results and conclude.

\section{Resonance model}
The production of $\phi$ mesons is described by a two step mechanism, 
i.e. the excitaton of nucleon resonances and their subsequent 
decay $pp\rightarrow pR \rightarrow pp\phi$.
The cross sections for the  resonance production were taken from ref.\cite{Teis} where
they were fitted to describe $\pi$, $\eta$, $\rho$, $\pi\pi$ production in
$NN$ collisions. The $\phi N R$ coupling is obtained from the 
known $\omega N R$ coupling of the 
$\omega$ meson \cite{Fuchs03} and the mixing angle between  
$\phi$ and $\omega$ mesons. The $\omega N R$ couplings, in turn, have
been determined within the framework of the extended Vector Meson Dominance
(eVMD)  model by fitting the available data on
electro- and photo- production of nucleon resonances as well as their
mesonic decays \cite{krivo02}. The description of the $\phi$ and $\omega$
meson production in elementary nucleon-nucleon reactions is then 
essentially parameter free since all model parameters have already 
been fixed by other sources. In \cite{Fuchs03} it was demonstrated 
that available data on the exclusive $\omega$ production in $pp$ reactions 
are very accurately reproduced by the present model over a wide 
energy range, i.e. from extremely close to threshold up to several 
GeV above threshold.  

The $pp\rightarrow pR \rightarrow pp\phi$ cross section is given as follows:

\begin{eqnarray}
\sigma (s)^{pp\rightarrow pp\phi}
=\sum_{R}\int_0^{(\sqrt{s}-2m_{p})^{2}}{d M^{2}}
\int_{(m_{p}+M)^{2}}^{(\sqrt{s}-m_{p})^{2}}d\mu ^{2}
\frac{ d\sigma (s,\mu )^{pp\rightarrow pR}}{d\mu ^{2}}
\frac{dB(\mu,M)^{R\rightarrow p\phi }}{d M^{2}}~~,
\label{sigNNV}
\end{eqnarray}
with $M$ being the running mass of $\phi$ meson.
The cross sections for the baryon resonances production are given by 
\begin{equation}
d\sigma (s,\mu )^{pp\rightarrow pR} = 
\frac{|{\cal M}_R|^2}{16 p_i \sqrt s \pi^2}\Phi_2(\sqrt {s},\mu,m_{p})dW_R(\mu)
\label{sigNR}
\end{equation}
with $\Phi_2(\sqrt {s},\mu,m_{p}) = \pi p^*(\sqrt {s},\mu,m_{p})/{\sqrt s}$ 
being the two-body phase space, 
$p^*(\sqrt {s},\mu,m_{p})$ the final c.m. momentum, $p_i$ the 
initial c.m. momentum.  $\mu$ and $m_R$ are the running and pole masses 
of the resonances, respectively, $m_p$  the proton mass. The
mass distribution $dW_{R}(\mu)$ of the resonances is described by the 
standard Breit-Wigner formula:
\begin{equation}
dW_{R}(\mu) = \frac{1}{\pi}\frac{\mu \Gamma_R (\mu)d\mu^2 }
{(\mu^2-m_{R}^2)^2 +(\mu\Gamma_R (\mu))^2}~.
\label{BW}
\end{equation}
The sum in (\ref{sigNNV}) runs over the 
same same set of nucleon resonances which is responsible for the 
$\omega$ meson production \cite{Fuchs03}. 
This includes all well established (4$*$) $N^*$ resonances quoted by the PDG
\cite{PDG} with masses below 2 GeV. The branching to the 
$\phi$ decay mode is given by 
\begin{equation}
dB(\mu,M)^{R\rightarrow p\phi } = 
\frac{\tan^2 \theta_{mix}\Gamma_{\rm N\omega}^R (\mu,M)}{\Gamma_R (\mu)}dW_{\phi}(M)~~,
\label{bra}
\end{equation}
with $\Gamma_{\rm N\omega}^R (\mu,M)$ calculated the same way as in the case of
$\omega$ meson production\cite{Fuchs03}.

The $\phi$ mass distribution $dW_{\phi}(M)$ is also described 
by a Breit-Wigner distribution, i.e. substituting $R \rightarrow \phi$ 
and $\mu \rightarrow M$ in Eq. (\ref{BW}). 
The energy dependence of the $\phi$ width 
$\Gamma_\phi (M)$ can be calculated according to

\begin{equation}
\Gamma_\phi (M) = \Gamma_\phi (m_\phi )\left(\frac{p^*(M,m_K,m_K)}{p^*(m_\phi ,m_K,m_K)}
\right)^3\left(\frac{m_\phi}{M}\right)^2~.
\label{phiwidth}
\end{equation}

Important properties of the $\Gamma_{\rm N\omega}^R(\mu,M)$ width
are the following ones: the $M$ dependence of magnetic, electric and 
Coulomb couplings entering into the amplitude and the Blatt-Weisskopf 
suppression factor which suppresses the width
for large off-shell masses $\mu$ of the resonances \cite{Fuchs03}. 
Their combined effects lead to an increase of the 
$\Gamma_{\rm N\omega}^R(\mu,M)$ width with $M$ increasing from $m_\omega$ to 
$m_\phi$ for $\mu > 2 GeV$. This effect is finally responsible for 
the violation of the naive OZI rule estimate (\ref{naive}).

The results for cross section $\sigma (pp\rightarrow pp\phi)$ are presented
in Fig. 1 in form of the ratio of the $\phi$ over $\omega$ meson production
cross sections. The $\omega$ mesons production cross section was shown
to agree well with experimental data if one uses a strong
$N^*(1535)N\omega$ coupling \cite{Fuchs03}
\footnote{Here we do not subract contributions from off-shell $\omega$ 
production from the total cross section since at the considered
energies, i.e.  
above the $\phi$ meson production threshold, these contributions 
are less than 10$\%$ of the total $\omega$ 
cross section. Hence the effect that can be approximatly neglected.}. 
The dashed line in 
Fig. 1 corresponds to the idealized case of stable
$\phi$ and $\omega$ mesons, i.e. to the limit of zero widths.

\section{Summary}
We have calculated the cross section for $\phi$ meson production in
$pp$ collision within the resonance model. The experimental data are well 
reproduced. The large violation of the OZI estimate 
for the $\phi$ over $\omega$ production, observed experimentally, 
is explained without introducing additional parameters into 
the present model. A further test of the resonance model 
could be provided by a comparison
of $\mu^2$ distributions (the distribution of the invariant mass of
proton-$\phi$ meson system) as soon as experimental data are available
for this observable.

One of us ( B.V.M.) is indebted to the Institute for Theoretical Physics of the
University of T\"{u}bingen for kind hospitality. This work is
supported  
by RFBR grant No.
03-02-04004, DFG grant No. 436 RUS 113/721/0-1. 


\begin{figure}[!htb]
\begin{center}
\includegraphics[width=.7\textwidth]{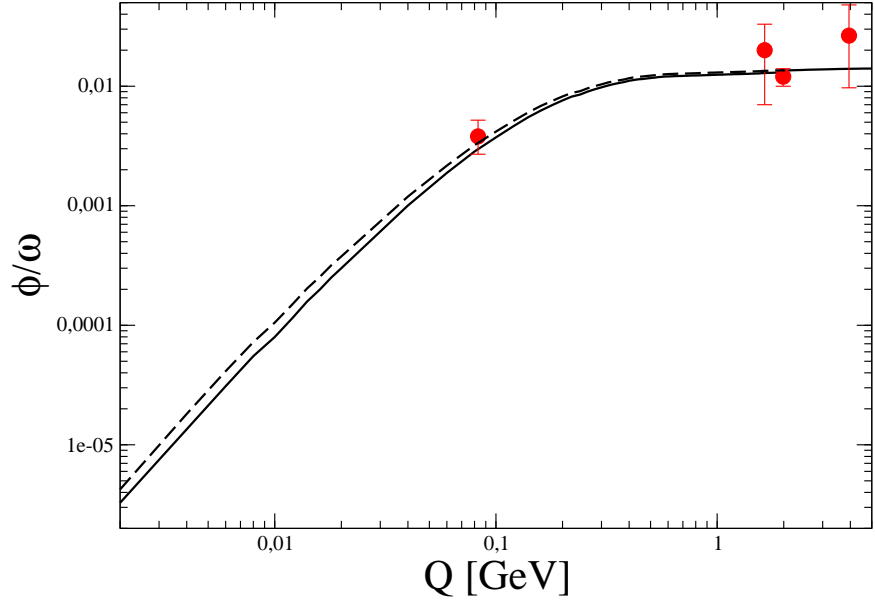}
\hspace{1cm}
\end{center}
\caption {The ratio of the $\sigma (pp\rightarrow pp\phi)$
over $\sigma (pp\rightarrow pp\omega)$  cross sections as a function of the c.m. energy
above the threshold $Q = \sqrt{s}-2m_p-m_\phi$ (solid line) is 
compared to experimental data \protect\cite{DISTO,Blo75,Bal77,Are82}. 
The dashed line corresponds to idealized case of stable
$\phi$ and $\omega$ mesons, i.e. the zero width limit.}
\end{figure}

\end{document}